\documentclass{article}

\usepackage{arxiv}

\usepackage[utf8]{inputenc} 
\usepackage[T1]{fontenc}    
\usepackage{hyperref}       
\usepackage{url}            
\usepackage{booktabs}       
\usepackage{amsfonts}       
\usepackage{amsmath}
\usepackage{nicefrac}       
\usepackage{microtype}      
\usepackage{lipsum}
\usepackage{graphicx}

\usepackage{diagbox}
\usepackage{caption}
\usepackage{subfigure}
\usepackage{floatrow}
\graphicspath{ {./images/} }

\title{Recurrent Feedback Improves Feedforward Representations in Deep Neural Networks}

\author{
  Siming Yan \\
  Computer Science Department \\
  Peking University\\
  \texttt{simingyanpku@gmail.com} \\
   \And
   Xuyang Fang \\
   Computer Science Department \\
   Carnegie Mellon University \\
   \texttt{xuyangf@andrew.cmu.edu} \\
   \AND
   Bowen Xiao \\
   Computer Science Department \\
   Peking University\\
   \texttt{mike.xiao@pku.edu.cn} \\
   \And
   Harold Rockwell \\
   Computer Science Department \\
   Carnegie Mellon University \\
   \texttt{hrockwel@andrew.cmu.edu} \\
   \AND
   Yimeng Zhang \\
   Computer Science Department \\
   and  Neuroscience Institute \\ 
   Carnegie Mellon University \\
   \texttt{yimengzh@andrew.cmu.edu} \\
   \And
   Tai Sing Lee \\
   Computer Science Department \\
   and  Neuroscience Institute \\
   Carnegie Mellon University \\
   \texttt{taislee@andrew.cmu.edu} \\
}

\DeclareMathOperator{\NNConv}{Conv}
\DeclareMathOperator{\NNDeconv}{Deconv}
\DeclareMathOperator{\NNSigmoid}{Sigmoid}
\DeclareMathOperator{\NNConcat}{Concat}
\DeclareMathOperator{\NNTanh}{Tanh}
\DeclareMathOperator{\NNPooling}{Pooling}

\begin{document}
\maketitle

\begin{abstract}
The abundant recurrent horizontal and feedback connections in the primate visual cortex are thought to play an important role in bringing global and semantic contextual information to early visual areas during perceptual inference, helping to resolve local ambiguity and fill in missing details. In this study, we find that introducing feedback loops and horizontal recurrent connections to a deep convolution neural network (VGG16)  allows the network to become more robust against noise and occlusion during inference, even in the initial feedforward pass. This suggests that recurrent feedback and contextual modulation transform the feedforward representations of the network  in a meaningful and interesting way. We study the population codes of neurons in the network, before and after learning with feedback, and find that learning with feedback yielded an increase in discriminability (measured by d-prime) between the different object classes in the population codes of the neurons in the feedforward path, even at the earliest layer that receives feedback.  We find that recurrent feedback, by injecting top-down semantic meaning to the population activities, helps the network  learn better feedforward paths to robustly map  noisy image patches to the latent representations corresponding to important visual concepts of each object class, resulting in greater robustness of the network against noises and occlusion as well as better fine-grained recognition. 



\end{abstract}

\section{Introduction}

The primate visual system is organized as a hierarchy of many visual areas with massive recurrent connections, both within each area and between different areas \cite{felleman1991distributed}. These recurrent connections  are thought to encode statistical priors of natural scenes, as well as to  bring in  higher order semantic and contextual information to help resolve local ambiguity, filling in missing details in lower visual areas during inference. There has been considerable recent interest in exploring and exploiting the use of recurrent feedback in deep convolutional neural networks \cite{li2018learning, nayebi2018task, jetley2018learn}.  
\begin{figure}[htb]
\centering
  \includegraphics[width=1\textwidth]{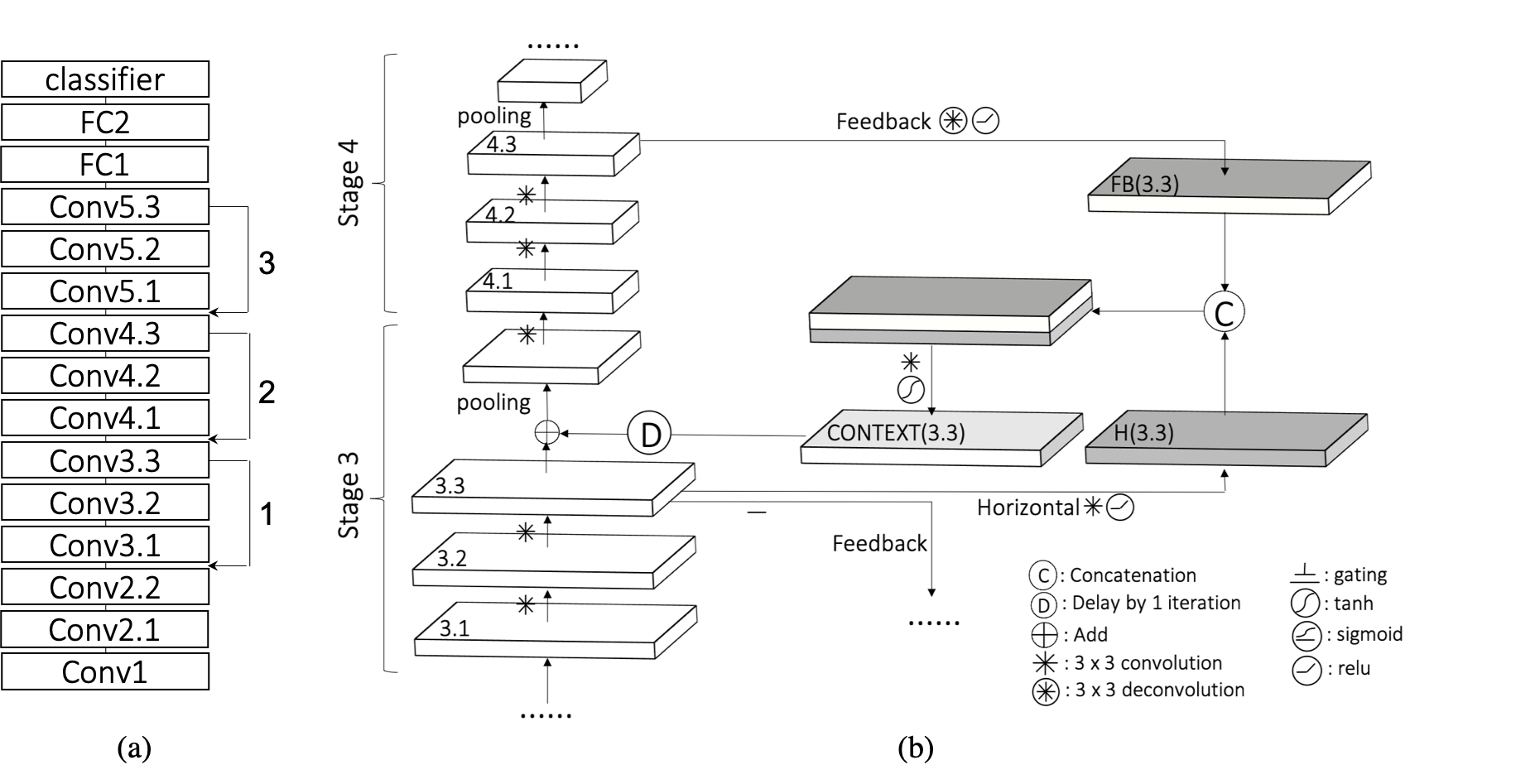}

  \caption{The schematic of a VGG Contextual Modulation(VGG-CM) model. (a) 3 loops between adjacent stages of VGG16. (b) microcircuit of recurrent interaction inside one loop. See Section~\ref{sub:feedback-loop}, ~\ref{sub:lateral} for details.}
  \label{fig:main_fig}
\end{figure}
This work shows promising results suggesting that recurrent connections allow networks achieve comparable performance in image classification with less parameters and fewer layers than a deep feedforward network.

These earlier studies focused on the contribution of recurrent connections for inference. Little is known about the impact of  recurrent connections on refining the feedforward computation. In this paper, we provide evidence that the recurrent feedback of semantic and contextual information  might highlight important visual concepts (clusters of latent representations, \cite{Wang2015Unsupervised}) at each level during training, steering  the feedforward circuits towards these beacons, resulting in a feedforward network that is more robust against noise, occlusion and adversarial attacks, and performs better in fine-grained recognition. 

\section{Related Work}

 Using global contextual information to disambiguate local early processing is a classic idea in computer vision \cite{marr1976cooperative, geman1987stochastic}. In this early work, recurrent connections were used to encode hand-crafted, generic statistical priors of natural scenes, such as smoothness and uniqueness constraints. This work was concerned primarily with early vision tasks, such as contour completion \cite{elder2003contour} and image reconstruction \cite{memisevic2010learning, mnih2010generating}. 
 
 Recently, there has been considerable interest in the computer vision community in exploring the functional advantages of top-down feedback and local recurrent connections in deep neural networks for object recognition  \cite{li2018learning,jetley2018learn, nayebi2018task, NIPS2018_8133}. \cite{li2018learning,jetley2018learn} used feedback essentially to introduce top-down information  from the fully semantic layers, enhancing relevant features in early layers in an attention-like mechanism. The large semantic gap between the semantic layer and early feature layers make feature alignment difficult, and such methods work mostly for low resolution CIFAR datasets  \cite{jetley2018learn}.  Other work has explored the use of local recurrence \cite{nayebi2018task, NIPS2018_8133} and/or long range loops \cite{nayebi2018task}, and was successful in demonstrating that local recurrent connections can achieve ImageNet object recognition performance that is comparable to a very deep Residual Network, but with much fewer layers and parameters. 
 
 A striking feature of the primate visual system is that the most massive recurrent feedback connections are between adjacent visual areas, and feedback can propagate all the way down to LGN. This ubiquitous feature has not been utilized in the above studies. The recurrent feedback in predictive coding network models \cite{NIPS2018_8133} updates the activities of neurons to reconstruct their input and is analogous to local recurrence within a visual area, rather than between adjacent visual areas. In this paper, we explore networks with multiple loops between adjacent stages of a deep convolution network, which is more consistent with the anatomical literature. We study how networks with local horizontal and top-down feedback can bring in contextual information to help resolve local ambiguity in  challenging scenarios. Our first contribution is in showing that VGG16, fine-tuned under recurrent feedback, becomes more robust against  noises, adversarial "attacks," and occlusion, and also performs better in fine-grained object recognition. Most interestingly, the lion's share of this improvement is due to the refinement in feedforward connections, not the effects of feedback during inference. Our second contribution is in characterizing the changes in latent representations of the networks underlying the improvement in performance. We are able to show that there is an increase in semantic clustering in the population activities subsequent to learning under feedback, and that feedback might have steered the network to learn more robust feedforward information processing paths  toward important clusters of latent representations, resulting in improvement in performance in challenging situations.

\section{Methods}


Our first objective is to characterize how contextual modulation created by local recurrent information from the current visual area  and top-down feedback from the adjacent higher visual area can be useful for improving network performance in object recognition.
Anatomical and neurophysiological evidence suggest that the surround contextual modulation a neuron experiences can be divided into a local component that is mediated by local circuits, and a global component that is mediated by top-down feedback \cite{angelucci2006contribution}. The near-surround includes surround with spatial extent that is 2-3 times the size of the receptive field, while the far-surround includes a space that is 3-6 times the size of the receptive field.

\subsection{Feedback loops -- contextual modulation from far surround}
\label{sub:feedback-loop}

We approximate the far-surround feedback architecture  by introducing feedback loops to adjacent stages in a standard VGG16 network \cite{simonyan2014very}.
We consider each stage in VGG, which contains multiple convolutional layers followed by a pooling layer as output, to be roughly equivalent to  a visual area. Thus, we model the feedback loop between adjacent visual areas by adding feedback loops between the last convolution layers before the pooling layer in each stage.  So, the three feedback loops added are from conv3.3 to conv2.2, from conv4.3 to conv3.3 and from conv5.3 to conv4.3 respectively (as shown in Figure 1a). 

Figure 1b illustrates the details of feedback loop 2 from conv4.3 (stage 4) to conv3.3 (stage 3). The feedback generates a tensor $\mathit{FB}$(3.3) with the same spatial resolution and feature dimensions as its target layer (i.e. conv 3.3)  by performing an expansion deconvolution using a 3 X 3 convolution filter on conv 4.3.
Since there are  three 3 x 3 convolutions and a pooling layer between conv 3.3 and conv 4.3,  each unit in the $\mathit{FB}$(3.3) layer has effectively integrated spatially global information from over 19 x 19 columns in conv 3.3, thus containing  information  coarser in spatial resolution but with higher order semantics, corresponding to the far-surround effect in neurophysiology \cite{angelucci2006contribution}.  The feedforward $R$ and the feedback $\mathit{FB}$ representations are specified by the following equations
\begin{align}
\label{eqs:feedforward}
R_{s(k)}^{t} & = \mathit{FF}_{s(k)}^{t} + \mathit{CM} \\
\mathit{FF}_{s(k)}^{t} & = \NNConv(\NNConv(\NNPooling(R_{s(k-1)}^{t}))) \\
\mathit{FB}_{s(k)}^{t} & = \NNSigmoid(\NNDeconv(R_{s(k+1)}^{t})) \end{align}
where  $R_{s(k)}$ are the unit responses  at time $t$ of  a particular VGG16 layer in the visual stage $k$, e.g. $s(2)$ stands for VGG16 conv 2.2, $s(3)$ for VGG16 conv 3.3, $s(4)$ for VGG16 conv 4.3, $s(5)$ for  VGG16 conv 5.3. These are the layers that provide as well as receive "inter-areal" feedback modulation. 
$R_{s(k)}^{t}$ is the sum of the feedforward input signal $\mathit{FF}_{s(k)}$ and a contextual modulation signal $CM$ to be described below. 
$\mathit{FB}_{s(k)}^{t}$ is  a tensor that contains the feedback information from the next stage $R_{s(k+1)}$, at the same spatial and feature dimension as $R_{s(k)}^{t}$  and is derived from expansion (up-sampling), followed by a 3 x 3 convolution. 

\subsection{Lateral connections - contextual modulation from near surround }
\label{sub:lateral}

\begin{figure}[htb]
\centering
  \includegraphics[width=\textwidth]{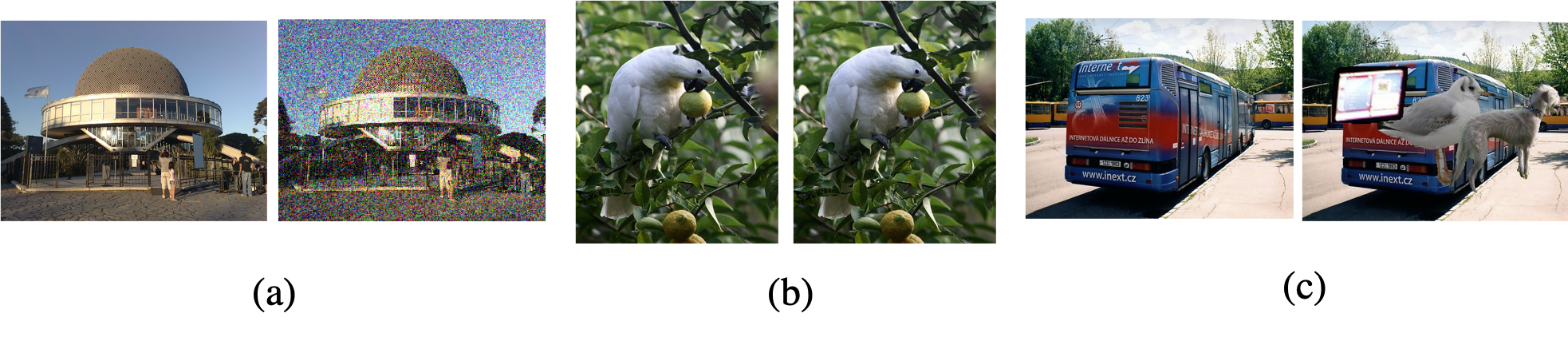}
  \caption{Examples of images used in three different challenging scenarios. Each pair contains the original image (left) and the corrupted image (right). (a): Corruption with 50\% Gaussian noise. (b) Corruption with 30\% adversarial attack noise. (c) Corruption with occlusion as in the VehicleOcclusion dataset \cite{wang2017detecting}.}
  \label{fig:examples}
\end{figure}
While top-down feedback provides semantic information of a more global nature, more detailed fine-grained feature constraints and priors are better encoded by local lateral or horizontal connections. For example, in V1, local connectivity has been shown to encode a contour continuation constraint  \cite{elder2003contour, li2006contour, iyer2017cortical} and surface smoothness constraint \cite{lee2008neural, koch1986analog}, providing a facilitatory associative field for contour and surface completion. Inhibitory local connectivity  can also implement a uniqueness constraint, mediating competitive surround suppression for redundancy reduction or sparse coding.   Here,  rather than specifying these  local priors explicitly by hand, as in traditional Markov random field models \cite{geman1987stochastic} or sparse coding models \cite{olshausen2004sparse, rozell2008sparse}, we used a convolutional layer ($H(3,3)$) to learn and model the local contextual priors implicitly using  3 x 3 convolution on the feedforward representational layer $R(3.3)$ as specified as follows. 
\begin{align}
\label{eqs:horizontal}
H_{s(k)}^{t} & = \NNConv(R_{s(k)}^{t})
\end{align}



\subsection{Contextual Modulation and Information Flow}
\label{sub:total_context}

The feedback tensor $FB_{s(k)}$ and the horizontal context tensor $H_{s(k)}$ are concatenated and then subject to 3 x 3 convolution to derive the contextual modulation tensor $\mathit{CONTEXT}_{s(k)}$ that is of the same dimension as as $R_{s(k)}$, which it modulates by point-wise addition.  The convolution is passed through a $tanh$ function to generate an output that can be positive or negative, as the contextual modulation effect can be inhibitory or excitatory, mediated by inhibitory and excitatory interneurons. 
\begin{align}
\label{eqs:contextual}
\mathit{CONTEXT}_{s(k)}^{t} & = \NNTanh(\NNConv(\NNConcat(H_{s(k)}^{t},\mathit{FB}_{s(k)}^{t}))) \\
R_{s(k)}^{t} & = \mathit{FF}_{s(k)}^{t} +  \mathit{CONTEXT}_{s(k)}^{t-D} 
\end{align}
where $D$ indicates a delay. For simplicity, we used a delay of $D=1$.


Given an input image, the network computes a first pass as a feedforward DCNN (unroll 0). At time $t=0$, there is no contextual modulation, i.e. CM = 0. For simplicity, we assumed there is no delay in computation through the entire feedforward hierarchy.  The context modulation tensors ($\mathit{CONTEXT}(2.2)$, $\mathit{CONTEXT}(3.3)$,$\mathit{CONTEXT}(4.3)$) are then computed based on feedback and horizontal integration and then added to the bottom-up activities of corresponding layers in the second pass (unroll 1), and continuing through the subsequent passes (unrolls). Because of the loops,  the bottom-up feedforward input $\mathit{FF}_{s(k)}$ activities of every layer will continue to be modified, and the activities of the higher layer will propagate down to modulate the activities in the lower layers, all the way down to conv2.2 in our current implementation.

\subsection{Training and Testing }

We train and test the networks on four benchmark datasets: CIFAR-10 (10 classes, low resolution 32 x 32 pixels images), CIFAR-100 (100 classes, low resolution 64 x 64 pixels images), ImageNet (1000 classes, high-resolution  224 $\times$224 pixels images) and CUB-200 (for fine-grained recognition). Our model is implemented in PyTorch. Training on the CIFAR and CUB-200 datasets is completed in 4 hours by 1 Nvidia GeForce GTX GPU, and on ImageNet in 2 days by 4 such GPUs. 

We then test the 3-loop network trained on ImageNet in three challenging scenarios: (1) Gaussian noise up to 50\% (meaning that the half-height bandwidth of the Gaussian is 50\% of the total range of image pixel values) added to the ImageNet images in the validation set; (2) Adversarial attack noise up to 30\% added to the ImageNet images; (3) Occlusion as in the VehicleOcclusion dataset \cite{wang2017detecting}. This occlusion dataset contains 4549 training images and 4507 testing images covering six types of vehicles, e.g., airplane, bicycle, bus, car, motorbike and train. For each test image in this dataset, some randomly-positioned, irrelevant occluders were placed onto the target object. Example images of each scenario are shown in  Figure~\ref{fig:examples}. 

In training the network, the hyperparameters were selected by grid search. For nonlinearity imposed after the convolution operation, we used ReLU for both integrating the feedback and the horizontal contextual information, as empirically we found them to work better than tanh and sigmoid.  We used tanh after the convolution on their concatenation C to generate the contextual modulation (CONTEXT) that can exert either positive (add) or negative (subtract) modulation to the feed-forward path.  Contextual modulation potentially can exert gain modulation (i.e. multiplication) on the feedforward path, rather than addition or subtraction. However, empirically we found that multiplication in this particular architecture actually produces inferior performance. We did not tried division, which might potentially model the divisive normalization mechanism in the cortex.

\begin{table*} [h]
\resizebox{\textwidth}{15mm}{%
  \centering
  \begin{tabular}{l|c|c|c|c|c|c|c}
  \toprule
  \diagbox{Models}{Experiments} & CIFAR-10 & CIFAR-100 & CUB-200 & ImageNet & Gaussian Noise 50\% & Adversarial Noise 30\% & Occlusion\\
  \midrule
  VGG16 & 91.20* & 67.06* & 64.880* & 71.076 & 12.983 & 42.541 & 34.500\\
  VGG-ATT & 91.77 & 69.48 & 73.200 & 71.213 & 13.002 & 42.759 & 37.619\\
  VGG-LR-2 & 91.49 & 68.99 & 72.990 & 71.551 & 14.041 & 43.623 & 38.721\\ 
  VGG-CM-0 & \textbf{92.10} & \textbf{71.51} & 74.824 & 71.645 & 17.970 & 47.249 & 50.012 \\
  VGG-CM-4 & 91.52 & 71.43 & \textbf{74.910} & \textbf{71.741} & \textbf{18.202} & \textbf{47.924} & \textbf{50.720} \\
  \bottomrule
  \end{tabular}
}{%
 \caption{Top-1 image classification accuracy on different experiments. Notice that in the CIFAR and CUB-200 experiments, VGG16 models have only one FC layer. VGG-ATT is the model proposed in \cite{jetley2018learn}, VGG-LR-n is the "rethinking" one-FC-layer VGG model with 2 unrolling times proposed in \cite{li2018learning} and tested with \textit{n} unrolling times. VGG-CM-\textit{n} is our 3-loop model trained with 4 unrolling times and tested with \textit{n} unrolling times.}
 \label{tab:cifar}
}
\end{table*}

\section{Experimental Results}

\subsection{Performance improvement in feedforward computation in noisy and occluded situations}

Table 1 provides a summary comparison between our VGG-CM with three loops against VGG16, as well as two other VGG-based models with recurrent feedback (VGG-ATT \cite{jetley2018learn} ; VGG-LR-2 \cite{li2018learning})  in their performance on four benchmark datasets (CIFAR10, CIFAR100, CUB200, and ImageNet), as well as three challenging test sets:  ImageNet images corrupted with Gaussian noise and adversarial noise, as well as occlusion. Our VGG-CM model's performance is comparable to VGG16 in CIFAR10 and ImageNet, but exhibits a significant improvement relative to VGG16 in fine-grained object recognition (15\% improvement), in noisy (30\% improvement for high level of Gaussian noises, and 13\% for adversary noise) and occlusion situations (47\% for VehicleOcclusion dataset). Our VGG-CM (with 3 loops and 4 unrolls) also outperformed  VGG-ATT and VGG-LR in these challenging noisy and occluded test sets.  The adversarial noise was generated by the standard Fast Gradient Signed Method (FGSM) attack.

\begin{table*}
  \centering
  \begin{tabular}{l|c|c|c|c|c|c}
  \toprule
  \cmidrule(r){1-2}
  \diagbox{Noise Level}{Models} & VGG16 & Unroll 0 & Unroll 1 & Unroll 2 & Unroll 3 & Unroll 4\\
  \midrule
  0 & 71.076 & 71.645 & 71.655 & 71.681 & 71.702 & 71.741 \\
  10 & 65.366 & 67.493 & 67.576 & 67.591 & 67.603 & 67.620 \\
  20 & 53.988 & 56.711 & 56.776 & 56.843 & 56.854 & 56.988 \\
  30 & 39.045 & 43.331 & 43.376 & 43.438 & 43.441 & 43.686 \\
  40 & 23.971 & 28.581 & 28.919 & 28.981 & 29.012 & 29.120 \\
  50 & 12.983 & 17.970 & 18.052 & 18.071 & 18.120 & 18.202 \\
  \bottomrule
  \end{tabular}
{
 \caption{Noisy image classification top-1 accuracy for different unrolling times of our proposed model. VGG16 is the standard feedforward VGG16 model and Unroll $n$ indicates $n$ unroll times during the test process.
}
 \label{tab:diff-unroll}
}
\end{table*}
  Table 2 shows the performance of VGG-CM at the different unrolls for different levels of Gaussian noise. Note that VGG-CM-0 is the VGG-CM at unroll 0, meaning that its performance is based purely on the first pass of the feed-forward computation, without any benefit of recurrent feedback contextual modulation. It is equivalent to removing all the loops and contextual modulation in our network in the testing stage, reducing to a network with exactly the same number of parameters as the original VGG16.
  Interestingly, VGG-CM-0's performance is very close to VGG-CM-4 in the four challenging tasks (fine-grained recognition, Gaussian noise, adversarial noise and occlusion), revealing that the majority of the performance improvement in challenging situations happens in the feedforward computation, with only small incremental benefits from the recurrent iteration during inference. More specifically, at 40\% noise level, unroll 0 achieves 89\% of the improvement attained by unroll 4; and at 50\% noise level, unroll 0 already achieves 96\% of the improvement attained by unroll 4. Thus, it seems that recurrent feedback and contextual modulation is very important during training, significantly modifying the feedforward connections of the system and accounting for the lion's share of the improvement, but actually plays a lesser role in our setup during inference, contributing only small incremental improvements.  
  
  Given the network showed most of the performance improvement in the feedforward connections and only minor improvement with additional recurrent iterations, i.e. unroll 4 is not significant better than unroll 0 or unroll 1, one wonder how many unrolls are necessary during training. Table 3 shows the test performance of the full network trained with different number of unrolls. The data suggests maybe one recurrent iteration during training is sufficient to produce most of the performance benefits. 
 
\begin{table*}
  \centering
  \begin{tabular}{l|c|c|c|c|c}
  \toprule
  \cmidrule(r){1-2}
  \diagbox{Noise Level}{Models} & VGG16 & Unroll 1 & Unroll 2 & Unroll 3 & Unroll 4\\
  \midrule
  0 & 71.076 & 71.132 & 71.313 & 71.316 & 71.741 \\
  10 & 65.366 & 66.468 & 66.484 & 67.481 & 67.620 \\
  20 & 53.988 & 55.980 & 55.938 & 56.894 & 56.988 \\
  30 & 39.045 & 42.392 & 42.516 & 43.551 & 43.686 \\
  40 & 23.971 & 28.110 & 28.544 & 29.031 & 29.120 \\
  50 & 12.983 & 17.941 & 17.954 & 17.982 & 18.202 \\
  \bottomrule
  \end{tabular}
{
 \caption{Object recognition accuracy at different noise levels for our full networks that are trained with different unrolls. Unroll 1 here means the network only has one iteration (unroll) during training.
}
 \label{tab:diff-unroll-train}
}
\end{table*}
  
  It should be noted that since VGG-CM-0, our network on the first pass, during inference or testing, has essentially the same number of parameters as VGG16, hence the improvement in performance is not simply due to the network being effectively deeper or having more parameters.
  The improvement in the robustness of the network thus primarily involves the refinement of the feedforward representations. As a side note, we found that having three loops is better than two loops and having two loops is better than one loop, even for unroll 0.  
  


\subsection{Increase in semantic and categorical discriminability of the feedforward latent representations}

What has happened to the feedforward latent representations that lead to the increase in the robustness of the network in these challenging situations? 
Our conjecture is that the cascade of feedback from  global and semantic representations of the higher layers has steered the representations at each level to become more semantically distinct. To evaluate this conjecture, we compute the average d-prime (or sensitivity index; distance divided by standard deviation) between all possible pairs of the 1000 object classes at each layer of latent representation along the feedforward hierarchy of VGG16 and VGG-CM-0.


Figures 3a and 3b show the average d-prime as well as the change in average d-prime between object classes for VGG-CM-0 and VGG-CM-4 relative to VGG16 at some selected layers of the network. They show that starting at conv2.2, the first target layer that receives the feedback,  VGG-CM's d-primes have substantially increased relative to that of the VGG16, supporting the idea that feedback has made the early representations  more semantically distinct at every layer. Like the robustness to noise, most of this improvement is in VGG-CM-0, showing that while this is learned with the help of feedback, the improvement itself no longer depends on feedback.


\begin{figure}
\centering
\begin{minipage}[c]{0.48\textwidth}
\centering
\includegraphics[width=6.5cm]{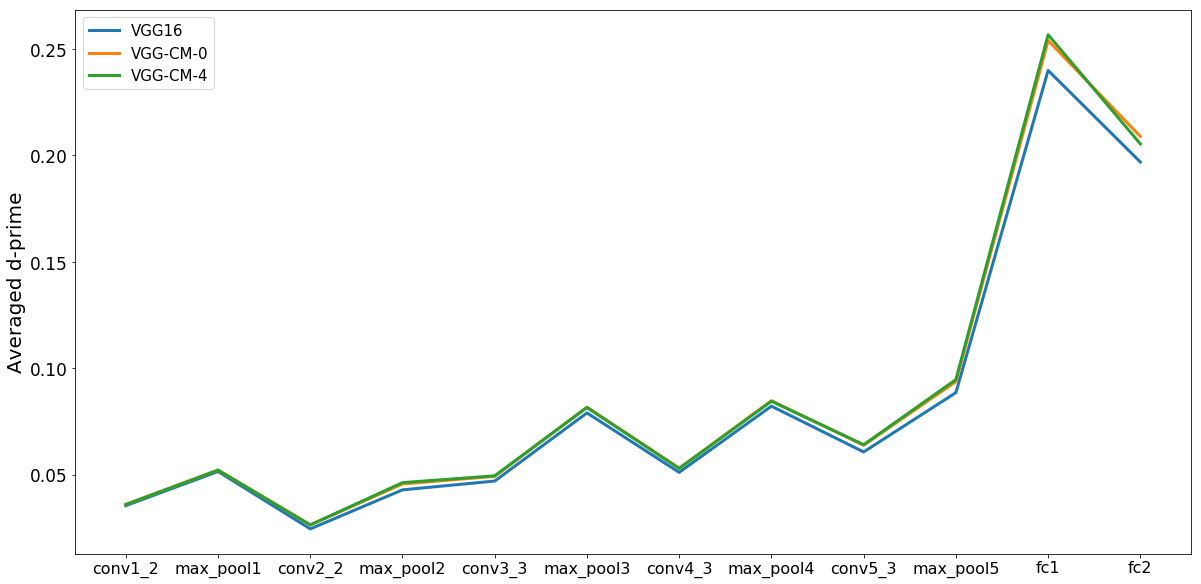}
\label{fig:sparsity}
\end{minipage}
\begin{minipage}[c]{0.5\textwidth}
\centering
\includegraphics[width=6.5cm]{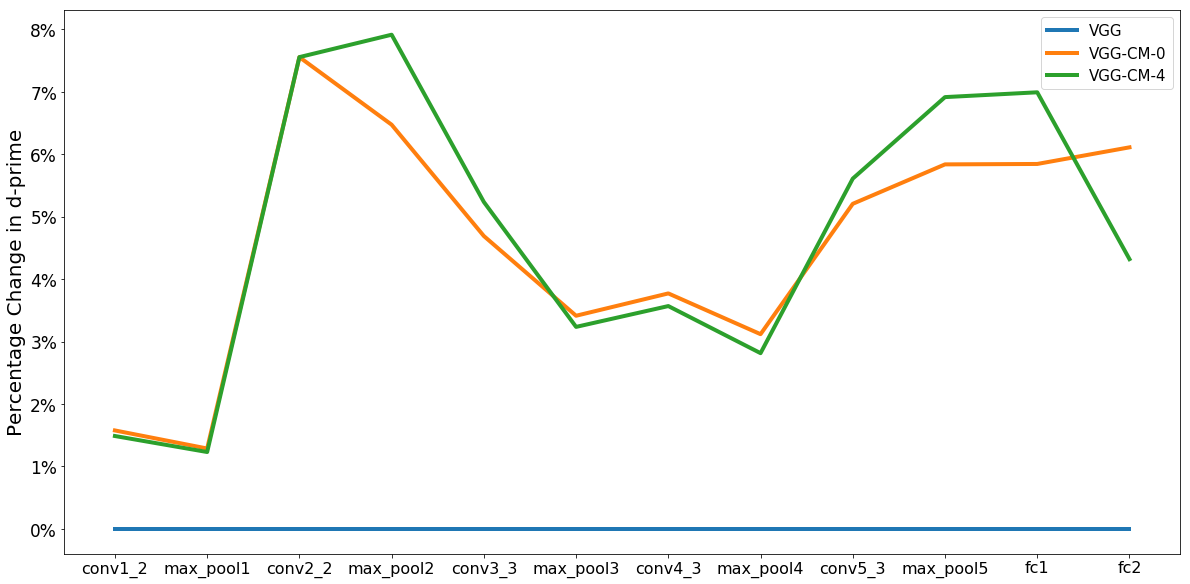}
\caption{(a) Average d-prime between all possible pairs of object classes for VGG-CM at unroll 0 as well as unroll 4 relative to VGG16. (b) Percentage changes in average d-prime relative to that of VGG for VGG-CM-0 and VGG-CM-4. }
\label{fig:d-prime}
\end{minipage}
\end{figure}

\subsection{Impact of feedback and horizontal modulation: ablation studies}

We performed a series of ablation experiments to find out whether top-down feedback alone, or horizontal recurrence alone is sufficient to generate the performance improvement. We found that combining both top-down feedback and horizontal connections yielded the best result in robustness against noise. An intuitive rationale for this is that top-down feedback introduces spatially more global surround but coarse contextual modulation, while horizontal recurrent connections bring in more precise local interaction, implementing more specific and precise local priors. Table 3 compares the performance of 
four models (standard VGG16, VGG-CM without feedback (no FB), VGG-CM without horizontal modulation (no H), VGG-CM full model) in different noise levels. 
For each version of the VGG-CM, we also compare the model's performance at unroll 0 (feedforward only) and at 4 unrolls (after 4 iterations of recurrent modulation). All the VGG-CM network models were trained with 4 unrolls.  The results show that horizontal modulation alone can bring about some improvement in the feedforward circuit, achieving 17\% of the 38\% increase achievable by the full model at high noise level. On the other hand, top-down modulation alone somehow cannot improve the feedforward circuit, though its' contribution is  important for the network to realize its full potential. It is possible that the top-down feedback is needed to gate the detailed computation by the horizontal connections that encode precise geometric and spatial priors as in the high-resolution buffer theory \cite{lee2003hierarchical},  predictive encoding model \cite{zhao2014predictive}, the gated Boltzmann machine \cite{memisevic2010learning}, or gated Markov random field \cite{mnih2010generating}. 
 
 \begin{table*}
  \resizebox{\textwidth}{15mm}{
  \begin{tabular}{l|c|c|c|c|c|c|c}
  \toprule
  \cmidrule(r){1-2}
  \diagbox{Noise Level}{Models} & VGG16 & No FB U0 & No FB U4 & No H U0 & No H U4 & Full U0 & Full U4 \\
  \midrule
  0  & 71.076 & 71.428 & 71.244 & 71.546 & 71.410 & 71.645 & 71.741 \\
  10 & 65.366 & 65.680 & 66.098 & 65.768 & 66.556 & 65.366 & 67.620 \\
  20 & 53.988 & 54.808 & 55.440 & 53.110 & 56.188 & 53.988 & 56.988 \\
  30 & 39.045 & 40.368 & 41.104 & 36.584 & 41.914 & 43.331 & 43.686 \\
  40 & 23.971 & 25.980 & 27.136 & 21.494 & 27.522 & 28.581 & 29.120 \\
  50 & 12.983 & 15.106 & 16.012 & 11.294 & 16.094 & 17.970 & 18.202 \\
  \bottomrule
  \end{tabular}
  }
{
 \caption{Object recognition accuracy at different noise levels, comparing four models -- standard VGG16, horizontal modulation only (No FB), top-down modulation only (No H), and both top-down and horizontal contextual modulation (Full). The models are compared at unroll 0 (U0, the feedforward case) and unroll 4 (U4, fully unrolled). 
}
 \label{tab:ablationtable}
}
\end{table*}
\vspace{-0.1in}

\subsection{Representational changes along the feedforward path? }

What might have happened to the feature representations to bring about this untangling of object representations along the hierarchy, even at the earlier layers? 
Given that the number of feature detectors at each layer is rather limited in the VGG, the changes in the receptive fields or feature tuning of the individual neurons are not obvious. We also looked for an increase in tuning selectivity, but found the changes in the population sparsity and life-time sparsity (tuning sharpness) of the neuronal populations to be negligible. The increase in semantic discriminability in the latent representations can arise from a increase in the distance in the population code or neural representational space between  the concepts or the reduction of the variability of the samples belong to a concept.


Next, we study the changes in the population codes by examining the changes in the visual concepts they encode. 
Visual concepts are  clusters of population activities associated with each class of visual objects that  correspond to parts of objects at different levels of scales and abstraction. For example, the face category will have visual concepts that correspond to the eyes and lips in lower layers, and  correspond to the a portion of the faces in the higher layers (see Figures 4a-c). Yuille and colleagues \cite{Wang2015Unsupervised}  have shown that k-means clusters of hypercolumn population activities  for each class of objects, with suitable pruning, give rise to semantically meaningful subparts and parts of objects, and that these parts can be used in a voting scheme to detect the existence of semantic parts, allowing the system to recognize an object based on its parts, under severe occlusion. For example, seeing an eye and a nose would be sufficient to recognize it is the face of a human.  They call these clusters "visual concepts". Figure 4a-c shows the clusters of visual concepts of face category (each row is a cluster, contains a number of examples) at three different layers for illustration. 

Given recurrent feedback has  made our VGG-CM more robust against noises and occlusion, we speculate that the feedforward path might have learned better visual concept representations. We applied the method described in \cite{Wang2015Unsupervised} to extract visual concepts from the max pooling layers of the third stage of the VGG network, for 99 of the 1000 ImageNet categories. This is accomplished by collecting the hypercolumn population response vectors across space and across images, and then perform K-means++ to cluster these population code, starting with a number of clusters matching the number of channels, which are then condensed via a greedy pruning method to yield the visual concepts.  

To assess whether there is any improvement in the visual concepts, we first measure the importance of each visual concept by the relative drop in object recognition performance the network experiences for that class of objects when that visual concept is removed from the intermediate representation at that layer. This is accomplished to air-brushing away the image patch in the  original image associated with that visual. We use the following score $S(V_m^k )$  to measure the importance of a visual concept $V_m^k$ for  category $m$,
\begin{equation}
        S(V_m^k)=\frac{\sum_{I_k^i \in M_k, P_{original}(I_k^i)>0.3}(\frac{P_{original}(I_k^i)-P_{occluded}(I_k^i)}{P_{original}(I_k^i)})}{|M|} 
\end{equation}
where $M_k \subset M$ are the images containing visual concept $V_m^k$, $P_{original} (I_k^i )$ is image $I_k^i$ ’s output probability for class $m$ in the classification layer of the neural network, and $P_{occluded} (I_k^i )$ is image $I_k^i$ ’s output probability after the visual concept $V_m^k$ is air-brushed away. The threshold of 0.3 for $P_{original} (I_k^i)$ is chosen to ensure that the difference is only measured for images that are reasonably well classified by the network in the first place. Specifically:
$$
P_{occluded}(I_k^i)=\frac{\sum_{P_{x,y}\in P_{k,l}^i} F(f_{l,x,y}^{air-brushed}) }{|P_{k,l}^i|}
$$
where $F(x)$ is the neural network's computation starting at layer $x$, and $f_{l,x,y}^{air-brushed}\in R^{W_l \times H_l \times N_l }$ is generated by air-brushing away a visual concept feature vector $p_{x,y} \in R^{N_l}$ from $I_{k,i}$'s intermediate layer response $f_l$. $P_{k,l}^i$ is the set of population responses in $f_l$ belonging to visual concept $V_m^k$. The airbrushing process point-multiplies the $5 \times 5$ spatial area around the visual concept with $1 - G(0,1)$, where $G(0,1)$ is a 2D Gaussian template with standard deviation 1, normalized to equal 1 in the center. In this way, the visual concept itself is entirely erased, and the nearby points that share its corresponding image patch are attenuated. The modified representation is then allowed to propagate through the remaining layers of the deep network to compute the probability of the different targeted classes. 

Note that the concepts are removed one at a time, so those that appear multiple times in an image (e.g. eyes) may have reduced importance by this measure. Figure 5 shows the ranked importance score thus computed of the 100 concept percentiles represented by the population codes of pool 3 layer for the  VGG16 network and the VGG-CM-0 network. The importance curve for 99 randomly selected classes of objects is rather sharp, indicating that a few visual concepts are significantly more important than the others for each class. 


\begin{figure}[]
\centering
  \includegraphics[width=5.5in]{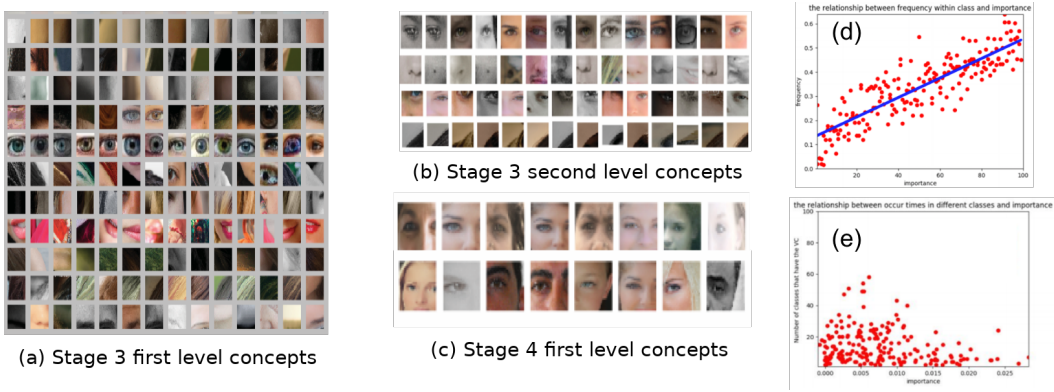}
  \caption{Some examples of visual concepts for faces at different stages, and plots of their within-class frequency and between-class uniqueness against their importance.}
  \label{fig:visual-concepst}
\end{figure}

What makes a visual concept important for object recognition? Intuitively, concepts (and their corresponding semantic parts) that appear frequently within one image class, but rarely within others, should be more useful for discriminating between those classes. We find this to be the case: within a class, the occurrence frequency of a concept is positively correlated with its importance (Figure 4d), and the number of similar concepts in different classes is negatively correlated with importance (Figure 4e). 

However, the ranked importance score curves of VGG16 and VGG-CM-0, as shown in Figure 5, are not different. This suggests that fine-tuning with recurrent feedback has not changed the visual concepts embedded in the population codes or, equivalently, that the robustness against noise and occlusion is not due to the visual concepts  becoming different, better or more important.  We next analyze what happened to the representations of these population codes when the images corresponding to the visual concept are corrupted with 50\% Gaussian noise. For each of the 99 chosen classes, we corrupt the images, pass them through the two networks, and calculate how many of the patches were still mapped into the same visual concept. We find that VGG-CM-0 performs better by this measure, retaining more of the noisy patches (Figure 5). Since the visual concepts have not fundamentally changed, mapping a noisy patch to the same population code as an unaffected patch should help the downstream neurons classify the concepts or object parts. We also compute the average distance of all the representations of the noisy samples relative to their original concepts' center, and found that indeed it is lower for the VGG-GM-0, meaning that the noisy population codes are not just more likely to be closest to, but are on average less deviated from their original visual concepts (Figure 5). Note that we only showed results from pool 3 concepts because these concepts tend to be semantically more meaningful than lower layer concepts and can be more reliably estimated than higher layer concepts \cite{wang2017detecting}.
These observations suggest that the recurrent feedback during training might have allowed higher level semantic and contextual information to highlight and select important visual concepts at each level, using these "beacons" to help the network to learn better paths to project more reliably to the same correct representations along the hierarchy. 

\begin{figure}[]
\centering
  \includegraphics[width=1.8in]{Figures/pool3plots_overlaid.png} 
  \includegraphics[width=1.8in]{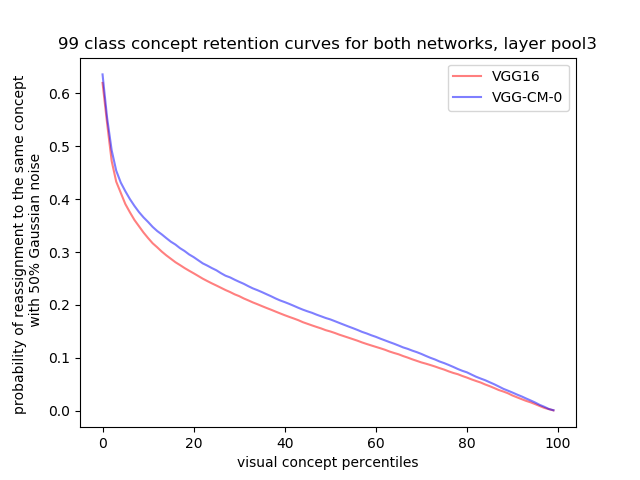} 
  \includegraphics[width=1.8in]{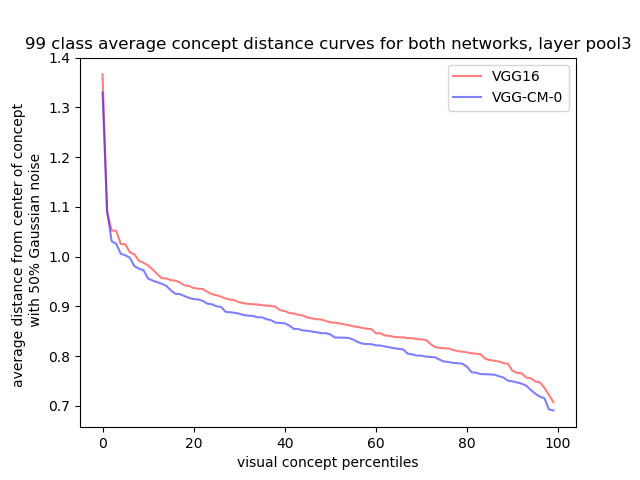}
  \caption{Plots of visual concept performance in VGG16 (red) and VGG-GCM-0 (blue), averaged over all 99 classes, ranked by percentile of the concepts. (left): The importance  of the visual concepts, as calculated by Equation (9). (center): The probability (or frequency) of reassignment of a patch to its original concept after the addition of 50\% Gaussian noise. (right): The average distance of feature responses from the cluster centers after the addition of 50\% Gaussian noise.}
  \label{fig:VC-difference}
\end{figure}

\section{Conclusion}
In this paper, we investigate a biologically-inspired neural architecture, with feedback loops between adjacent stages of processing as well as horizontal contextual modulation in a deep convolutional neural network. We find that fine-tuning such a network with recurrent feedback and contextual modulation allows it to become more robust against noise and occlusion, as well as perform better at fine-grained discrimination. We develop an approach to evaluate the latent representations of deep neural networks based on the notion of visual concepts. We find that while the visual concepts in the feedforward latent representations have not been changed by feedback, the domain of images mapped to these visual concepts have been increased, allowing corrupted versions of the image patches to be mapped to the representation of the semantically meaningful visual concepts. Our finding suggests that higher order semantic feedback might have highlighted the visual concepts at the lower level during training, providing  beacons to help the network find its way to the appropriate target concepts, thus improving its robustness. 


\bibliographystyle{unsrt}  
\bibliography{references}  

\end{document}